\documentstyle[11pt]{article}
\begin{document}
\def\beq{\begin{eqnarray}}
\def\eeq{\end{eqnarray}}
\def\IM{{\rm {Im}}}
\def\RM{{\rm {Re}}}
\def\nn{\nonumber}
\def\Sin{{\rm sin}}
\def\Cos{{\rm cos}}
\def\PRD{Phys. Rev. D}
\def\PRL{Phys. Rev. Lett}
\def\PLB{Phys. Lett. B}
\def\EPJC{Euro. Phys. J. C}

\centerline{\large\bf Exploring CPT violation in $B\to
J/\psi K$ decays using kaon regeneration}

\vspace{2cm}
 \centerline{ Zhengtao Wei }

\vspace{0.8cm}

\begin{center}
CCAST(World Laboratory), ~~P.O.Box $8730$, Beijing
$100080$, China,

Institute of Theoretical Physics, Academia Sinica, P.O.Box
$2735$, Beijing $100080$, China \footnote{mailing address}

\end{center}

\vspace*{0.5cm}

\begin{center}\begin{minipage}{12cm}

{\bf Abstract}

We present an analysis of CPT violation in $B\to J/\psi K$
decays. A method of kaon regeneration is proposed to
increase the $K_{S(L)}$ interference effects. With the time
evolution of both neutral B and K mesons, we show that it
is possible to determine the CPT violating parameter in
$B^0-\bar{B^0}$ mixing or constrain it.

\end{minipage}
\end{center}

\vspace{1.5cm}
 PACS number: 11.30.-j, 13.25.-k.

Key words: CPT violation, cascade mixing, kaon
regeneration.

\newpage

\baselineskip 24pt


With the running of two B factories and the proceeding of
the future B physics projects ( BTeV and LHCb etc.), a
large number of B mesons will be accumulated to test the
standard model and explore new physics beyond it. The CPT
invariance is an important theorem of local, relativistic
field theory, but its validity must be checked by precise
experiment. Except kaon decays, B decays can provide another
ideal place to perform such test. The exploration of CPT
violation in B decays has been got many theoretical
interests recently [1-8]. In experiment, only OPAL
collaborations measured the the CPT violation in neutral B
meson oscillation \cite{OPAL}.

Besides the uncertainties in the standard model, CPT
violation is a new physics effect and may be entangled with
other new physics effects. In \cite{Xing}, authors point
out that the CPT violation violation and the violation of
$\Delta B=\Delta Q$ rule can not be distinguished by the
opposite-sign dilepton asymmetry of neutral B decays. These
new physics effects increase the difficulty of extracting
the information of CPT violation. The study of CPT
violation is unavoidable to consider some complex systems.

$B\to J/\psi K$ decay is an interesting decay process. It
is well known that $B\to J/\psi K_S$ decay is a "gold-
plated" mode to determine clean $sin 2\beta$ through time
dependent rate asymmetries between $B^0$ and $\bar B^0$.
Recent preliminary results from the BABAR and BELLE
collaborations stimulates theoretical interests about this
decay. Another feature of $B\to J/\psi K_S$ decay is
"cascade decay". In the decay chain $B\to J/\psi K\to
J/\psi [f]_K$ where the initial $B^0-\bar B^0$ mixing is
followed by neutral K mixing, this "cascade mixing" process
contains more information than the ordinary decays. Azimov
\cite{Azimov} first pointed out this unique feature. In
\cite{Kayser}, Kayser shows that the interference of
$K_{S(L)}\to \pi l\nu$ can be used to explore $cos 2\beta$.
But this method suffers from the small branching ratio of
$K_S\to \pi l\nu$. One procedure to solve this problem is
proposed in \cite{Quinn} by using kaon regeneration. This
procedure utilizes the fact that kaon mass eigenstates are
not same in matter and in vacuum. With the kaon
regeneration, the interference between the $K_S$ and $K_L$
decays to $\pi \pi$ provides information to determine $cos
2\beta$. Here we want to point out this interference
provides enough knowledge to extract the information of CPT
violation.

Kaon regeration is an important experimental tool to study
the discrete symmetry violation. The direct CP violation
$\RM (\epsilon'/\epsilon)$ is measured through
the kaon regeneration. In \cite{Wei}, cascade mixing is
used to explore CPT violation where CPT violation comes
from the $B^0-\bar B^0$ mixing. This study discuss the kaon
decays in vacuum and has the disadvantage discussed above.
In this paper, our purpose is to use the kaon regeneration
to discuss the CPT violation in cascade decay $B\to J/\psi
K\to J/\psi [2\pi]_K$.

Let us first discuss $B^0-\bar{B^0}$ mixing including CPT
violation. The oscillation of  $B^0$ and $\bar{B^0}$ caused
by weak interaction leads to the mass eigenstates are not
the flavor states but their superpositions. For neutral
$B^0$ system, the two eigenstates can be generally given by
\begin{eqnarray}
|B_1>=\frac{1}{\sqrt{|p_1^2|+|q_1|^2}}[p_1|B^0>+q_1|\bar{B^0}>]
\nonumber\\
|B_2>=\frac{1}{\sqrt{|p_2^2|+|q_2|^2}}[p_2|B^0>-q_2|\bar{B^0}>]
\end{eqnarray}
and their eigenvalues are
\begin{eqnarray}
\mu_1=m_B-\frac{\Delta
m_B}{2}-\frac{i}{2}(\Gamma_B+\frac{\Delta \Gamma_B}{2})
     =m_B-\frac{i}{2}\Gamma_B-\frac{\Delta m_B}{2}-\frac{i}{2}y\Gamma_B
\nonumber\\ \mu_2=m_B+\frac{\Delta
m_B}{2}-\frac{i}{2}(\Gamma_B-\frac{\Delta \Gamma_B}{2})
     =m_B-\frac{i}{2}\Gamma_B+\frac{\Delta m_B}{2}+\frac{i}{2}y\Gamma_B
\end{eqnarray}
The quantities $p_i$, $q_i$ are mixing parameters. In the
standard model, mixing parameters $p_i$, $q_i$ are usually
parameterized by small quantities $\epsilon$, $\delta$
which represents CP, T violation and CPT violation
respectively. As we have known from the standard model, CP
violation in B decays is expected to be large. So it is
convenient to introduce the exponential parameterization
for $p_i$, $q_i$ as
\begin{eqnarray}
\frac{q_1}{p_1}= tg\frac{\theta}{2}e^{i\phi}, ~~~~
\frac{q_2}{p_2}=ctg\frac{\theta}{2}e^{i\phi}
\end{eqnarray}
where $\theta$ and $\phi$ are complex phases in general. In
the standard model, $\theta=0$, $\phi\approx -2\beta$.

Define
$(\frac{q}{p})_B=\sqrt{\frac{q_1}{p_1}\frac{q_2}{p_2}}$ and
CPT violating parameter $\theta'\equiv
\theta-\frac{\pi}{2}$, then $\theta'\neq 0$ represents
indirect CPT violation in $B^0-\bar{B^0}$ mixing.

From Eq.(1) and Eq.(3), one can obtain the evolution of the
initially pure $B^0$ or $\bar{B^0}$ state after proper time
$t$ as
\begin{eqnarray}
|B^0(t)>=g_+(t)|B^0>+\bar g_+(t)|\bar{B^0}> \nonumber\\
|\bar B^0(t)>=g_-(t)|\bar{B^0}>+\bar g_-(t)|B^0>
\end{eqnarray}
where
\begin{eqnarray}
g_+(t)=f_+(t)+cos\theta f_-(t), ~~~&g_-(t)=f_+(t)-cos\theta
f_-(t) \nonumber\\ \bar{g}_+(t)=sin\theta e^{i\phi}f_-(t),
~~~~& \bar{g}_-(t)=sin\theta e^{-i\phi}f_-(t)
\end{eqnarray}
and
\begin{eqnarray}
f_+(t)=\frac{1}{2}(e^{-i\mu_L t}+e^{-i\mu_H t})
  =e^{-im_B t-\frac{1}{2}\Gamma_B t}ch(\frac{ix-y}{2}\Gamma_Bt)
\nonumber\\ f_-(t)=\frac{1}{2}(e^{-i\mu_L t}-e^{-i\mu_H t})
  =e^{-im_B t-\frac{1}{2}\Gamma_B t}sh(\frac{ix-y}{2}\Gamma_Bt)
\end{eqnarray}

Next we turn to the $K^0-\bar{K^0}$ mixing. Up to now, the
experiment in neutral K systems have gained a high level
and find no new physics effects. Moreover, it is difficult
to explore new physics effects of K system from B decays.
So it is reasonable to neglect the new physics effects of K
system when studying B decays. We take the formulae for
$K^0-\bar{K^0}$ mixing within the standard model. The
eigenstates of the neutral K mesons can be represented by
\begin{eqnarray}
|K_S>=\frac{1}{\sqrt{|p_K^2|+|q_K|^2}}[p_K|K^0>+q_K|\bar{K^0}>]
\nonumber\\
|K_L>=\frac{1}{\sqrt{|p_K^2|+|q_K|^2}}[p_K|K^0>-q_K|\bar{K^0}>]
\end{eqnarray}
and their eigenvalues are
\begin{eqnarray}
\mu_{S(L)}=m_K{\stackrel{(+)}{-}}\frac{\Delta m_K}{2}
           -i\frac{\Gamma_{S(L)}}{2}
\end{eqnarray}
where $m_K$ is the average of the $K_S$ and $K_L$ masses,
$\Gamma_{S,L}$ are the $K_{S,L}$ widths.

The time evolution of $K_{S(L)}$ state in vacuum obeys the
simple exponential law \beq K_S(t)=e^{-i\mu_S t}K_S(0),
~~~~~~ K_L(t)=e^{-i\mu_L t}K_L(0) \eeq Due to the
scattering of $K^0$ and $\bar{K^0}$ with the nuclei in the
matter, the eigenstates are not $K_{S(L)}$ but their
mixture. The relation between them is: $K_S'\sim K_S-rK_L$
and $K_L'\sim K_L+rK_S$, where $r$ is the regeneration
parameter. The difference of the eigenstates $K_S'$, $K_L'$
in matter and the eigenstates $K_S$, $K_L$ in vacuum can be
used as an effective tool to convert the coherent mixture
of $K_S$ and $K_L$. We use the regeneration formulae given
in \cite{Quinn}. Considering neutral kaon hits one
regenerator with length $L$ at time $t=0$, after time $t$,
it passes through the regenerator. The kaon components
$\alpha_{S(L)}$ in nautral kaons is written as \cite{Quinn}
\beq \alpha_i(t)=e^{-N(\sigma_T+\bar{\sigma_T})L/4}
            e^{-i\mu_S t}\sum_{j=S,L}m_{ij}\alpha_j(0)
\eeq where $\sigma_T(\bar{\sigma_T})$ is the total cross
section of $K^0(\bar{K^0})$ forward scattering, the
regeneration parameters $m_{ij}$ is taken to linear order
in $r$, $m_{SS}\sim 1$, $m_{SL}\sim m_{LS}\sim
r[e^{-i(\mu_L-\mu_S)t}-1]$, and $m_{LL}\sim
e^{-i(\mu_L-\mu_S)t}$.

Now consider the decay chain $B\to J/\psi K\to J/\psi f$
where $f$ refers to $2\pi$ in our paper. At time $t_B$, the
neutral B decays to $J/\psi K$. Then after time $t_1$, the
neutral kaons hits a kaon regenerator. At time $t_2$, the
kaon passes through the regenerator and again evolves in
vacuum until at time $t_K$ it decays to $f_K$. In the standard
model, $B^0$ can only decay to $J/\psi K^0$ while
$\bar{B^0}$ decays to $J/\psi \bar{K^0}$. The decay of
$B^0\to J/\psi \bar{K^0}$ and $\bar B^0\to J/\psi K^0$
occurs due to higher order effects which are estimated to
be very small and negligible. The possible new physics
effects can cause $B^0\to J/\psi \bar{K^0}$ and $\bar
{B^0}\to J/\psi K^0$ and these effects are similar to the
violation of $\Delta B=\Delta Q$ rule. When considering the
CPT violation, these effects should be considered. Define
the nonrephasing variables \beq
 \lambda &\equiv& (\frac{q}{p})_B \frac{\bar {A_{\bar f}}}
  {A_f}(\frac{p}{q})_K, \nn\\
 y' &\equiv& \frac{A_{\bar f}}{A_f}(\frac{p}{q})_K, \nn\\
 \bar y' &\equiv& (\frac{q}{p})_B\frac{\bar A_f}{A_f}
\eeq
where
\beq
 A_f=A(B^0\to J/\psi K^0), ~~~~~
 A_{\bar f}=A(B^0\to J/\psi \bar{K^0}), \nonumber\\
 \bar{A_f}=A(\bar{B^0}\to J/\psi K^0), ~~~~~
 \bar{A_{\bar f}}=A(\bar{B^0}\to J/\psi \bar{K^0}).
\eeq

The components of $K_{S(L)}$ in $B^0{\stackrel{t_B}{\to}}
J/\psi K$ decay are
\beq
 \alpha_S=\frac{\sqrt{p_K^2+q_K^2}}{2p_K}A_f
            [f_+(t_B)+\lambda f_-(t_B)-\theta'f_-(t_B)
            +y'f_+(t_B) +\bar y'f_-(t_B)]\nn\\
 \alpha_L=\frac{\sqrt{p_K^2+q_K^2}}{2p_K}A_f
            [f_+(t_B)-\lambda f_-(t_B)-\theta'f_-(t_B)
            -y'f_+(t_B)+\bar y'f_-(t_B)]
\eeq

So the decay rate of $B^0{\stackrel{t_B}{\to}}J/\psi K
{\stackrel{t_K}{\to}}J/\psi f_K$ is
\beq
 \Gamma(B^0{\stackrel{t_B}{\to}}J/\psi K
    {\stackrel{t_K}{\to}}J/\psi f_K)=
    \frac{1}{2}e^{-\Gamma_B t_B}
    \Gamma(B^0\to J/\psi K^0)\nn\\
    \times e^{-N(\sigma_T+\bar{\sigma_T})L/2}
    e^{-\Gamma_St_K}\Gamma(K_S\to f)~~~~~~~~~~~~~~~~\nn \\
    \times\{|a_{SS}+\eta a_{LS}|^2|\alpha_S|^2
    +|a_{SL}+\eta a_{LL}|^2|\alpha_L|^2 ~~~~~~~~\nn \\
    +2\IM[(a_{SS}+\eta a_{LS})(a_{SL}+\eta a_{LL})^*]
    \IM(\alpha_S\alpha_L^{*})~~\nn \\
    +2\RM[(a_{SS}+\eta a_{LS})(a_{SL}+\eta a_{LL})^*]
    \RM(\alpha_S\alpha_L^{*})
    \}
\eeq where $\eta=\frac{A(K_L\to f_K)}{A(K_S\to f_K)}$,~
$a_{SS}=m_{SS}$,~
$a_{SL}=m_{SL}e^{-i(\lambda_L-\lambda_S)t_1}$,~
$a_{LS}=m_{LS}e^{-i(\lambda_L-\lambda_S)(t_K-t_2)}$,~
$a_{LL}=m_{LL}e^{-i(\lambda_L-\lambda_S)t_1}$.

The decay rate of $\bar{B^0}{\stackrel{t_B}{\to}}J/\psi K
{\stackrel{t_K}{\to}}J/\psi f_K$ can be similarly obtained.
The four terms in Eq.(14) corresponding to different passes
from the initial B meson to the final state. The first and
second terms in the first line arise from $B^0 \to J/\psi
K_S$ and $B^0\to J/\psi K_L$ respectively. The second and
third lines arise from the interference of $B^0 \to J/\psi
K_{S(L)}\to J/\psi f_K$. In the standard model, the second line
gives the information of $\Cos2\beta$. This is the Kayser's
original proposal to determine $\Cos2\beta$ \cite{Kayser}.

From Eq.(14), one can see that cascade decay has rich time
evolution behavior. The time behaviors of neutral kaon
decay are contained in the coefficients of $a_{ij}$ which
can be determined from the experiment. $\alpha_S$ and
$\alpha_L$ represents the time behavior of neural B meson
decay. Their explicit formulae with new physics of CPT
violation are listed in the Appendix. Without new physics,
this formulae will go back to the form within the standard
model \cite{Quinn}.

From the known parameter of neutral K system, we can fit
the unknown parameter in B system. There are 8 unknown
parameters in total: the real and imaginary part of
$\lambda, \theta', y', \bar y'$ in total. Combining decay
rate of ${\stackrel{(-)}{B^0}}\to J/\psi K\to J/\psi f_K$
with kaon evolves in matter and in vacuum, we have enough
information to determine these parameters in principle.
However, it is difficult to determine all the unknown
parameters in practice. If the small $y'$ and $\bar y'$ can
be neglected, only the unknown $\lambda$ and $\theta'$ are
left. The multiparameter fit can be used to determine them.
At least experiment can give a constraint on CPT violating
parameter $\theta'$.

In our proposal, the B meson corresponds the uncorrelated case
which B meson is tagged by its flavor. Our method of
extracting the CPT violating parameter $\theta'$ can be
extended to the correlated case. In such case, the odd
function of B meson decay time $\Sin\Delta m_Bt_B$ term can
be cancelled by the decay rate asymmetry \cite{Wei}. This can lead to
a further simplification.

In summary, it is possible to extract the CPT violation in
$B\to J/\psi K$ decays using kaon regeneration.



\section*{Appendix}
Only considering the new physics of CPT violation
\beq
 |\alpha_S|^2&=&\Cos^2\frac{\Delta m_Bt_B}{2}
  +|\lambda|^2\Sin^2\frac{\Delta m_Bt_B}{2}
  +(\IM\theta'-\IM\lambda)\Sin\Delta m_Bt_B\nn\\
  &&-(\RM\theta'\RM\lambda+\IM\theta'\IM\lambda)
  (1-\Cos\Delta m_Bt_B) \nn\\
 |\alpha_L|^2&=&\Cos^2\frac{\Delta m_Bt_B}{2}
  +|\lambda|^2\Sin^2\frac{\Delta m_Bt_B}{2}
  +(\IM\theta'+\IM\lambda)\Sin\Delta m_Bt_B\nn\\
  &&+(\RM\theta'\RM\lambda+\IM\theta'\IM\lambda)
  (1-\Cos\Delta m_Bt_B) \\
 \IM(\alpha_S\alpha_L^*)&=&\RM\lambda\Sin\Delta m_Bt_B
  +(\IM\theta'\RM\lambda-\RM\theta'\IM\lambda)
  (1-\Cos\Delta m_Bt_B)\nn\\
 \RM(\alpha_S\alpha_L^*)&=&\Cos^2\frac{\Delta m_Bt_B}{2}
  -|\lambda|^2\Sin^2\frac{\Delta m_Bt_B}{2}
  +\IM\theta'\Sin\Delta m_Bt_B\nn
\eeq


\end{document}